\documentclass[reprint,amsmath,amssymb,aip,jmp]{revtex4-1} 	

\usepackage{bmpsize}
\usepackage{graphicx,color}								
\usepackage{amssymb}
\usepackage{natbib}
\usepackage{array}
\usepackage{subfigure}

\begin{document}

\title{The nature of geometric frustration in the Kob-Andersen mixture}

\author{Peter Crowther}
\affiliation{H.H. Wills Physics Laboratory, Tyndall Avenue, Bristol, UK}
\affiliation{School of Chemistry, University of Bristol, Cantock's Close, Bristol, UK}
\author{Francesco Turci}
\email[Corresponding author: ]{f.turci@bristol.ac.uk}
\affiliation{H.H. Wills Physics Laboratory, Tyndall Avenue, Bristol, UK}
\author{C. Patrick Royall}
\affiliation{H.H. Wills Physics Laboratory, Tyndall Avenue, Bristol, UK}
\affiliation{School of Chemistry, University of Bristol, Cantock's Close, Bristol, UK}
\affiliation{Centre for Nanoscience and Quantum Information, Tyndall Avenue, Bristol, UK}

\begin{abstract}
Geometric frustration is an approach to the glass transition based upon the consideration of locally favoured structures (LFS), which are geometric motifs which minimise the local free energy. Geometric frustration proposes that a transition to a crystalline state is frustrated because these LFS do not tile space. However, this concept is based on icosahedra which are not always the LFS for a given system. The LFS of the popular Kob-Andersen (KA) model glassformer is the bicapped square antiprism, which does tile space. Such an LFS-crystal is indeed realised in the $\mathrm{Al_{2}Cu}$ structure, which is predicted to be a low energy state for the KA model with a 2:1 composition. We therefore hypothesise that upon changing the composition in the KA model towards 2:1, geometric frustration may be progressively relieved, leading to larger and larger domains of LFS which would ultimately correspond to the $\mathrm{Al_{2}Cu}$ crystal. Remarkably, rather than an increase, upon changing composition we find a small decrease in the LFS population, and the system remains impervious to nucleation of LFS crystals. We suggest that this may be related to the \emph{composition} of the LFS, as only a limited subset are compatible with the crystal. We further demonstrate that the $\mathrm{Al_{2}Cu}$ crystal will grow from a seed in the KA model with 2:1 composition and identify the melting temperature to be 0.447(2).
\end{abstract}

\maketitle

\section {Introduction}
Understanding the glass transition remains one of the major outstanding questions of condensed matter physics~\cite{cavagna2009, berthier2011}. Broadly speaking, theories can be classified into those which propose a transition to an ``ideal glass'' state at finite temperature~\cite{adam1965, lubchenko2007, parisi2010} and those which propose that there is no thermodynamic transition but that structural relaxation becomes progressively slower upon cooling towards absolute zero~\cite{tarjus2005, chandler2010}. Since at some point a glass forming liquid cannot be equilibrated on an experimental timescale, it is hard to establish whether its structural relaxation time $\tau_\alpha$ would diverge at finite temperature (in the case of an ``ideal glass'') or whether it it would remain finite (although large) until temperature drops to absolute zero. Fits to experimental data are insufficient to discriminate either scenario conclusively~\cite{hecksler2008, mauro2009}.

Although the absolute nature of the glass transition remains unclear, in recent years, structural change has been measured in a number of materials approaching dynamical arrest~\cite{royall2015physrep}. Structural change is necessary in the case that there is a thermodynamic transition, but it is not excluded by the no-transition scenario. This structural change is hard to detect in two-point correlations such as the radial distribution function $g(r)$ but can be detected using higher-order contributions such as three-body correlations~\cite{dicicco2003, coslovich2013}. Other methods include identifying geometric motifs which minimise the local free energy, called locally favoured structures~\cite{royall2015physrep, frank1952, leocmach2013} and order-agnostic methods where the nature of the structural correlations are not specified~\cite{kurchan2009, mosayebi2010, cammarota2012pnas, cammarota2012epl, sausset2011, dunleavy2012}.

The theory of geometric frustration makes explicit reference to locally favoured structures (LFS)~\cite{tarjus2005}. In d=3 the archetypal LFS is the icosahedron identified by Frank when considering the monatomic Lennard-Jones model~\cite{frank1952}. Icosahedra do not tile Euclidean space, thus inducing frustration and preventing crystallisation. The situation is rather different in curved space where 600 perfect (strain-free) tetrahedra comprising 120 particles can be embedded on the surface of a four-dimensional hypersphere~\cite{coexeter, coexeterpolytope, sadoc1981, nelson, nelson1983, straley1984}. Each particle in this 4d Platonic solid or ``polytope'' is at the centre of a 12-particle icosahedrally coordinated shell. Mapping from curved space back to Euclidean space, introduces defects in the tiling of icosahedra which leads to frustration and the avoidance of a transition to a fully icosahedral state. Thus the degree of deviation in curvature from that of the fully icosahedral state corresponds to the strength of frustration. For small deviations, frustration is weak and large domains of icosahedra can form. Further deviation in the curvature corresponds to strong frustration and smaller domains of icosahedra.

Locally favoured structures other than icosahedra are also possible. In d=2, the LFS is often a hexagon which does tessellate in 2d space. This means that curving space induces frustration rather than relieving it as in d=3~\cite{modes2007,sausset2008, sausset2010, tarjus2010, sausset2010pre}. It has been shown that LFS domain sizes can be very large in the case of weak frustration; furthermore, the degree of curvature controls the domain size as predicted by geometric frustration theory~\cite{sausset2010,sausset2010pre}. Again, in d=3 the LFS need not be the icosahedron~\cite{royall2015physrep} but may be other motifs depending which system is considered~\cite{coslovich2007, hocky2014, royall2014}. A number of investigations have been made into LFS domain size in d=3, with most~\cite{charbonneau2012, charbonneau2013pre, charbonneau2013jcp, royall2014, dunleavy2014, malins2013fara, malins2013jcp} (but not all~\cite{tanaka2010, kawasaki2010jpcm, leocmach2012}) finding that the characteristic lengthscale of the domains is small. This implies that in 3-dimensional Euclidean space, geometric frustration is strong~\cite{royall2014}.

The idea of curving space to relieve geometric frustration is predicated upon the LFS being icosahedral~\cite{coexeter, coexeterpolytope, sadoc1981, nelson, tarjus2005}. In general this is not the only possible method of relieving frustration. Examples include the use of polydispersity to control frustration in systems where the locally favoured structure was crystalline~\cite{tanaka2010, kawasaki2010jpcm} though others found that the LFS in weakly polydisperse hard spheres was based on five-membered rings with strong geometric frustration~\cite{charbonneau2012, royall2014, dunleavy2014}. A number of glassformers have been found to crystallise, often into complex crystal structures which incorporate the LFS ~\cite{pedersen2010}. Demixing into simpler crystals such as FCC is also possible~\cite{punnathanam2006}, as is the formation of substitutional FCC lattices~\cite{jungblut2012}

\begin{figure}[t]
\centering
\includegraphics[scale=1]{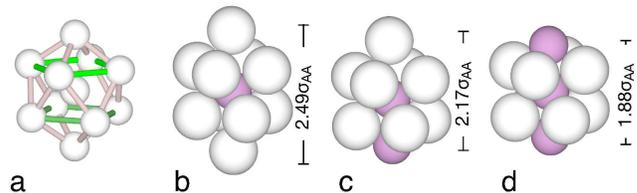}
\caption{\label{figKAtypes} (a) The geometric form of a bicapped square antiprism, the LFS for the Kob-Andersen binary mixture. (b-c-d) The three most common configurations of the LFS in a KA liquid with the larger A particles in white and the smaller B particles in purple: (b) $\mathrm{A_{10}B_{1}}$, (c)$\mathrm{A_{9}B_{2}}$ and (d)$\mathrm{A_{8}B_{3}}$. Only the $\mathrm{A_{8}B_{3}}$ form tessellates to form a bulk $\mathrm{Al_{2}Cu}$ crystal. Centre-to-centre distances between the particles defining the principal rotational axis of energy-minimised clusters are reported next to each bicapped square antiprism's type.}
\end{figure}

In this work we consider the Kob-Andersen (KA) binary Lennard-Jones model, a popular 3d model glassformer in which the LFS is a bicapped square antiprism of 11 particles [see Fig.~\ref{figKAtypes} (a)]~\cite{coslovich2007, malins2013fara}. We believe that the structure of the LFS is related to the non-additive interactions between the two species in the KA model. Supporting this hypothesis, certain metallic glasses with non-additive interactions also exhibit considerable populations of bicapped square antiprisms~\cite{evteev2003,royall2015physrep,cheng2011}. The structure of the LFS in the KA model is interesting in the context of geometric frustration because it tessellates in Euclidean space into a crystal of the form $\mathrm{Al_{2}Cu}$~\cite{fernandez2003pre}. Although the full phase diagram for the KA model has yet to be determined, some state points have been considered. The 1:1 mixture crystallises readily to a CsCl structure~\cite{fernandez2003pre,bannerjee2013} while the 4:1 composition is predicted to phase separate into coexisting face centred cubic lattices~\cite{toxvaerd2009}. Evidence in support of these scenarios has been found upon quenching, but around the 2:1 composition no evidence of either FCC or CsCl was found~\cite{valdes2009}. Here we primarily consider the 2:1 composition and in particular the role of the $\mathrm{Al_{2}Cu}$ structure. Previous work with this composition found some change in local geometry in the supercooled liquid ~\cite{fernandez2004} but focused on smaller structures than the bicapped square antiprism which was later identified as the LFS~\cite{coslovich2007, malins2013fara}.

The $\mathrm{Al_{2}Cu}$ crystal offers a new means to control geometric frustration in the Kob-Andersen model. Rather than curvature, the composition of the mixture can be modified between the 4:1 of the normal mixture to 2:1, and the size of the LFS domains can be measured to determine the degree of frustration. We begin by considering the normal 4:1 mixture, measuring the LFS population and investigating the domain size. Comparing this to the 3:1 and 2:1 compositions for a given degree of supercooling, we unexpectedly find a small decrease in the LFS population as the composition is changed, rather than an increase. We will show that this appears to be related to the need to consider the composition of the LFS rather than just the number of LFS in the system. In moving from the 4:1 composition to the 2:1 composition, we find a significant increase in the population of LFS which are compatible with the $\mathrm{Al_{2}Cu}$ crystal, but this remains a fraction of the total LFS population. We find no evidence for nucleation of the $\mathrm{Al_{2}Cu}$ crystal but we are able to grow it from a seed and estimate the melting temperature as 0.447(2).

This paper is organised as follows. After describing our simulation and analysis methodology in section~\ref{sectionMethods}, we present details of the dynamical behaviour of our simulations in section~\ref{sectionDynamics}. We then show that the composition of the LFS is altered by modifying the composition of the KA model in section~\ref{sectionLocallyFavouredStructures}. In section~\ref{sectionCrystallisation} we proceed to show that the 2:1 composition can undergo seeded crystal growth in MD simulations to give a crystal of the form $\mathrm{Al_{2}Cu}$. We discuss and interpret our findings in section~\ref{sectionDiscussion} and conclude in section~\ref{sectionConclusion}.

\section{Methods}
\label{sectionMethods}

\subsection{Model}

The classical Kob Andersen (KA) binary Lennard-Jones mixture is a model glassformer composed of 80\% large particles (A) and 20\% smaller particles (B) and is designed to be homologous to the metal alloy, $\mathrm{Ni_{80}P_{20}}$, a metallic glass\cite{weber1985, kob1995a}. The Lennard-Jones parameters are, $\sigma_{AA}=\sigma$, $\sigma_{AB}~=~0.8\sigma$, $\sigma_{BB}~=~0.88\sigma$, $\varepsilon_{AA}~=~\varepsilon$, $\varepsilon_{AB}~=~1.5\varepsilon$ and $\varepsilon_{BB}~=~0.5\varepsilon$. The particles have the same mass, $m_{A}~=~m_{B}$. The interactions are truncated at $r_{c}^{AA}~=~2.5\sigma$, $r_{c}^{AB}~=~2.0\sigma$ and $r_{c}^{BB}~=~2.2\sigma$.

In the present work, we consider the usual 4:1 composition as well as two additional mixtures with the same interactions but modified ratios of A to B particles (3:1 and 2:1). The units used throughout this work are reduced Lennard-Jones units, with respect to the A particles. This means that length is measured in units of $\sigma$, energy (E) in units of $\varepsilon$, density ($\rho$) in units of $N\sigma^{3}/V$, time (t) in units of $\sqrt{m\sigma^{2}/\varepsilon}$, pressure (P) in units of $\varepsilon/\sigma^{3}$ and the Boltzmann constant ($k_{b}$) is set to 1. Throughout we used the LAMMPS molecular dynamics package~\cite{Plimpton1995} and set pressure $P=0$.

\subsection{Structural Analysis}
In order to detect locally favoured structures in the considered systems, we employ the Topological Cluster Classification (TCC) algorithm. The TCC is described in detail elsewhere~\cite{malins2013tcc} but in summary, it identifies a neighbour network for all of the particles in the system using a Voronoi tessellation and then determines shortest path relations for 3, 4 and 5 membered rings. From these basic structures, more complex structures are identified in a hierarchical fashion. The structure that we are interested in identifying is the LFS, the bicapped square antiprism. In the language of the Voronoi Face Analysis~\cite{tanemura1977} this corresponds to the (0,2,8) class~\cite{coslovich2007}. Because we are considering a binary mixture, the composition of the LFS may be significant. The TCC is able to detect the composition of structures that it recognises in terms of the number of ``A'' and ``B'' particles that make up the cluster. We find that there are three main LFS compositions. All have a small B particle in the centre and differ in whether the other two positions along the principal rotational axis (or spindle) of the structure are occupied by A or B particles (see Fig.~\ref{figKAtypes}). There is some small variation in structure, with a B particle occasionally occupying a non-spindle position, though the three structures shown account for 95\% of the LFS in a 2:1 KA liquid at T = 0.44. We refer to these three structures as the $\mathrm{A_{10}B_{1}}$, the $\mathrm{A_{9}B_{2}}$ and the $\mathrm{A_{8}B_{3}}$ clusters.

\section{Results}
\label{sectionResults}

\subsection{Slow Dynamics}
\label{sectionDynamics}
In order to characterise the relevant time scales of the KA mixtures, we have determined their relaxation times over a range of temperatures (see Fig.~\ref{figAngell}a). We performed simulations of $N=3375$ particles at $P=0$ in the isobaric-isothermal ensemble for three different mixtures: the 2:1, 3:1 and 4:1 compositions. 
After a quench from a random (infinite temperature) configuration the systems were equilibrated at each temperature and from these equilibrium simulations, the intermediate scattering functions were computed, providing an estimate for the $\alpha$-relaxation times $\tau_{\alpha}$. For each temperature, the equilibration time is at least 100$\tau_{\alpha}$.

The non-Arrhenius growth of the relaxation times as a function of the inverse temperature can be fitted with the conventional Vogel-Fulcher-Tamman (VFT) equation

\begin{equation}
\tau_{\alpha}(T)=\tau_{\infty}\exp\left[\frac{DT_0}{(T-T_0)}\right]
\label{eq:VFT}
\end{equation}

\noindent which allows for the computation of the fragility $D$ and the ideal-glass transition temperature $T_0$, at which the relaxation times appear to diverge. These values are reported in Table~\ref{vftTable} and show that the change in composition leads to a significant increase in the transition temperature $T_0$ when moving from the 4:1 to the 2:1 composition accompanied by an increase in fragility. Without entering into debate about the physical significance of the VFT fit~\cite{cavagna2009}, we use $T_0$ to rescale our data for convenient comparison of the degree of supercooling between the different compositions. As discussed in the following section, this change in composition is related to important modifications in the structure of the liquids. 

\begingroup
 \squeezetable
\begin{table}[h]
\begin{ruledtabular}
\begin{tabular}{llll}
VFT  & 4:1   & 3:1   & 2:1   \\ \hline
$\mathrm{T_{0}}$   & 0.278 & 0.300 & 0.336 \\ 
$\mathrm{\tau_{\infty}}$ & 0.122 & 0.132 & 0.142 \\
D    & 2.81  & 2.46  & 2.10  \\
\end{tabular}
\end{ruledtabular}
\caption{Parameters of VFT fits to relaxation times for KA liquids of different compositions.}
\label{vftTable}
\end{table}
\endgroup

\subsection{Locally Favoured Structures}

\label{sectionLocallyFavouredStructures}
\begin{figure*}[t]
\centerline{\includegraphics[scale=1]{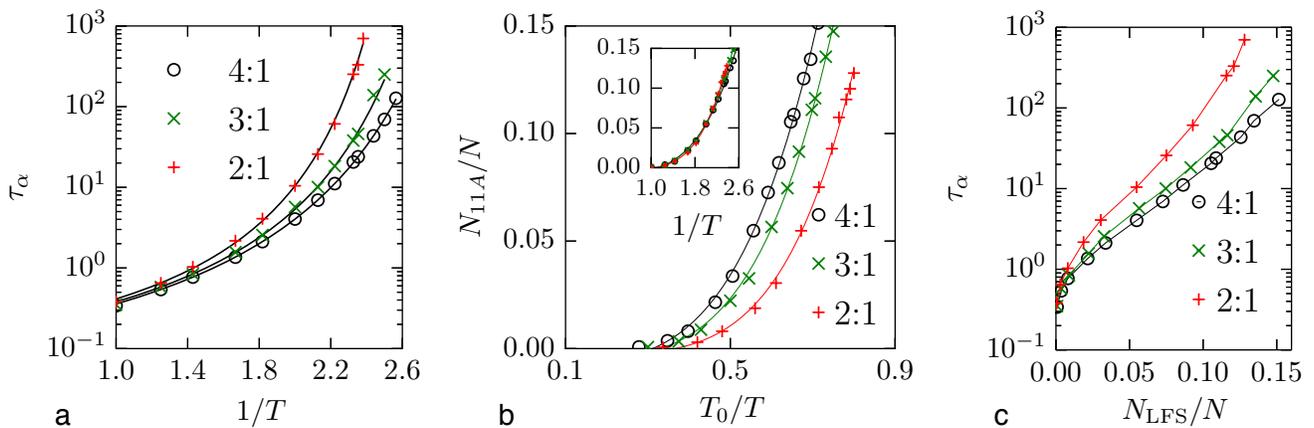}}
\caption{\label{figAngell}
(colour online) (a)
Angell plot for KA liquids with different ratios of large to small particles at P = 0. Black lines indicate fits according to Eq.~\ref{eq:VFT}. (b) The fraction of particles that are in a cluster of LFS for different ratios of large to small particles as a function of the degree of supercooling $T_0/T$ where $T_0$ is reported in Table~\ref{vftTable}. In the inset, the same fractions plotted against the reduced temperature, with no rescaling. The relaxation times and the fraction of LFS are compared in panel (c), where an important slowing down is observed for the 2:1 composition.}
\end{figure*}

In simulations of the KA mixture, the population of LFS is inversely related to temperature~\cite{coslovich2007, malins2013fara}. Fig.~\ref{figAngell}(b) shows the proportion of particles in a KA liquid that are in a cluster of LFS at $P=0$ for the three mixtures considered in this study.

We observe that the three mixtures have similar LFS concentrations and that, upon rescaling the temperature with respect to the fitted ideal glass transition temperature $T_{0}$, the 2:1 mixture shows \emph{fewer} structures for a given degree of supercooling [see inset of Fig.~\ref{figAngell}(b)]. The expected trend is that modifying the system such that the LFS is compatible with the crystal should increase the population of bicapped square antiprisms but our results are contrary to this.

We can consider the relation between structure and dynamics by observing the relation between the LFS concentration and the relaxation time of the system, see Fig.~\ref{figAngell}c. In this case, it is possible to observe the slowing of the dynamics induced by the change in the stoichiometry. For an LFS concentration of approximately 10\% the relaxation time of the 2:1 mixture exceeds that of the 4:1 mixture by an order of magnitude. Therefore, it appears that the slowing dynamics of the system is correlated to a change in the composition of the locally favoured structures, and in particular to an increase in the proportion of $\mathrm{A_{8}B_{3}}$ clusters.

\begin{figure}[h]
\centerline{\includegraphics[scale=1]{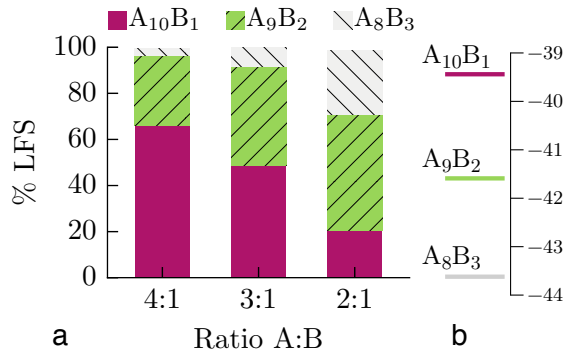}}
\caption{\label{figKARatiosGraph}
(colour online) (a) Composition of LFS clusters in KA mixtures of varying A:B particle ratios at $T=1.41T_{0}$. (b) Energies of isolated LFS clusters of different compositions, in reduced units.} 
\end{figure}

\begin{figure*}[tb]
\centerline{\includegraphics[scale=1]{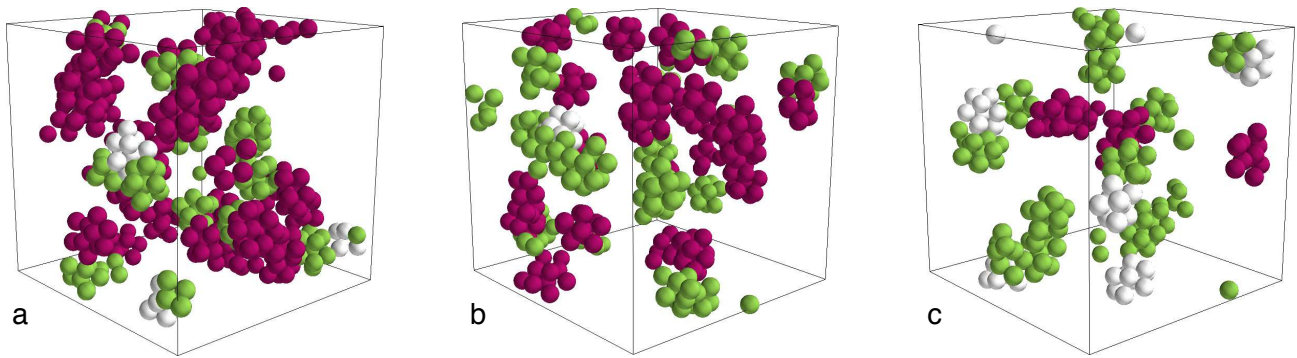}}
\caption{\label{figFluid11As}
(colour online) Network of LFS for (a) 4:1, (b) 3:1 and (c) 2:1 Kob-Andersen liquids at $T= 1.41 T_0$. Purple particles indicate clusters of stoichiometry $\mathrm{A_{10}B_{1}}$, green are $\mathrm{A_{9}B_{2}}$ and white are $\mathrm{A_{8}B_{3}}$.}
\end{figure*}

Figure~\ref{figKARatiosGraph}(a) shows the variation in LFS structure for the three different KA mixtures that we have considered at a temperature $T = 1.41T_0$. Next to it, in Fig.~\ref{figKARatiosGraph}b, we report the ground state potential energies of the same clusters, evaluated in isolated configurations using an efficient basin-hopping method~\cite{GMIN,wales1997}. The composition that best minimizes energy in the bicapped square antprism is the $\mathrm{A_{8}B_{3}}$ cluster. We note that it is this cluster which is compatible with the AlCl$_2$ crystal, the other two are not. Despite its low energy, the $\mathrm{A_{8}B_{3}}$ cluster is almost absent in the classical 4:1 mixture in fact, which is dominated by the highest energy cluster $\mathrm{A_{10}B_{1}}$. Increasing the proportion of B particles in the mixture through the 3:1 to the 2:1 mixture leads to a significant increase in the relative amount of $\mathrm{A_{8}B_{3}}$ to ($\approx25\%$) of the LFS. Although these energetically optimal $\mathrm{A_{8}B_{3}}$ clusters do tessellate, as demonstrated in Fig.~\ref{figFluid11As}, they are never clustered together in the fluid mixtures and during our simulations never attain a critical size sufficient to form a crystalline nucleus. This aside, they do optimize the local energy landscape and their presence in the 2:1 mixture correlates with the exceptional slowing down observed in Fig.~\ref{figAngell}c.

\begin{figure}[h]
\centering
\includegraphics[scale=1]{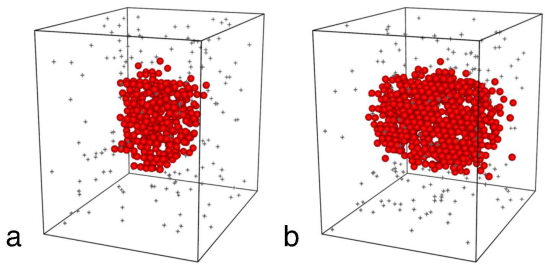}
\caption{\label{FigSnaps} (colour online) Snapshots of the system at different times: (a) $T=0.425$, immediately post quench and (b) $T=0.425$, after 500$\tau_{\alpha}$. Centres of crystalline LFS are represented in red, centres of non-crystalline LFS are represented by crosses.}
\end{figure}

\subsection{Crystallisation}
\label{sectionCrystallisation}

Given that we see no crystalline order or even significant increases in the size of connected LFS domains supercooled liquids, we enquire as to the stability of the $\mathrm{Al_{2}Cu}$ crystal. Motivated by the work of Fern\'{a}ndez and Harrowell~\cite{fernandez2003pre} who considered the crystallisation of a CsCl crystal from a KA mixture using a crystal seed, we conducted seeded crystal growth molecular dynamics simulations in the isobaric-isothermal ensemble at P~=~0 with a 2:1 ratio KA mixture.

First we determined the density of the bulk crystal in the $\rm Al_2Cu$ conformation at zero pressure and constant temperature for all the low temperatures studied in a system of 3375 particles. Then, simulations were performed by taking a crystal of $\mathrm{Al_{2}Cu}$ consisting of 20736 particles at the previously determined density and fixing a 1500 particle cubic region in the centre of the simulation box. The surrounding particles were equilibrated at $T=2$ and then quenched to low temperatures. In this way we approximate a 2:1 KA liquid surrounding an ideal $\mathrm{Al_{2}Cu}$ seed. The central seed was then unconstrained and the growth process tracked in the isothermal-isobaric ensemble for a time equivalent to 500$\tau_{\alpha}$ , where the relaxation time refers to the corresponding 2:1 KA liquid at P~=~0 (see section~\ref{sectionDynamics}).

To determine the size of the crystalline region, we coarse-grain by considering only the centres of the bicapped square antiprisms. Using this coarse-graining, we define domains of LFS using an agglomerative single linkage hierarchical clustering method~\cite{jain1988}, identifying the largest domain as the growing crystal.

In this way, we detect the stable crystalline structure from the LFS belonging to the liquid region, as shown in the two snapshots in Fig.~\ref{FigSnaps}. From direct inspection of the configurations, we notice that the crystal grows in a cylindrical fashion, with the main axis aligned with the primary rotational symmetry axis of the bicapped square antiprisms. Few new clusters are added to the lateral surface of the cylinder throughout the simulations, suggesting a strong anisotropy in the surface tensions of the different crystalline orientations. 

\begin{figure}[t]
\centerline{\includegraphics[scale=1]{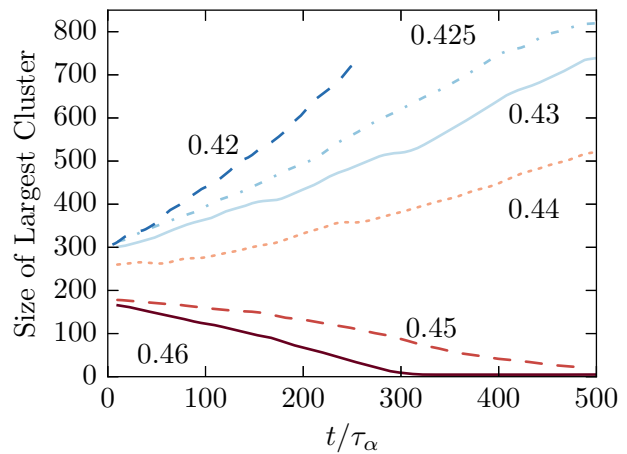}}
\caption{\label{figcrystalgrowth}
Time evolution of the number of bicapped square antiprism centres in the largest crystalline cluster for different temperatures. Time $t=0$ corresponds to the instant at which the initially frozen crystalline nucleus is freed and begins to thermalize. Each temperature is an average of 5 simulations. }
\end{figure}

We considered this system for a range of temperatures $T=0.42-0.46$ which spanned from crystal growth to the dissolution of the seed into the melt. The simulation was performed 5 times at each temperature with randomised initial particle velocities, the average crystal growth is plotted in Fig.~\ref{figcrystalgrowth}. From this series of non-equilibrium simulations we infer the temperature $T_{\rm melt}$ at which the growth rate is zero.

\begin{figure}[t]
\centerline{\includegraphics[scale=1]{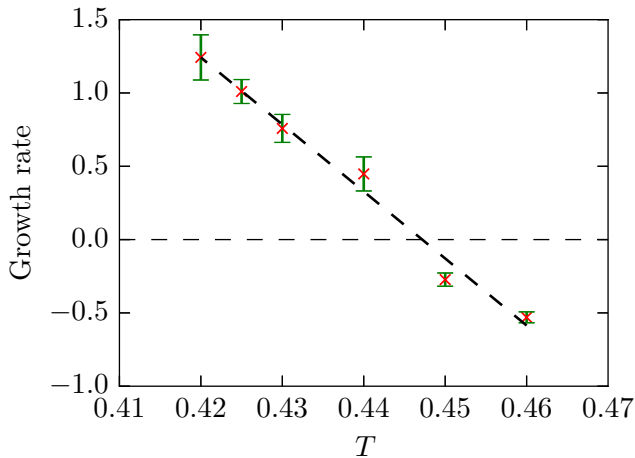}}
\caption{\label{figCrystalGrowthRate}
Growth rate of the 2:1 KA crystal as a function of the temperature, measured by linear fits to the growth curves in Fig.~\ref{figcrystalgrowth}. The dashed line is a linear fit to these points. The crossing point between the linear fit and the zero-line delivers an estimate for the melting temperature $T_{\rm melt}\approx0.447(2)$.}
\end{figure}

For temperatures $T\leq0.45$ we observe the melting of the initially seeded structure (see Fig.~\ref{figcrystalgrowth}). This allows us to estimate positive and negative growth rates via linear fits to the growth curves (Fig.~\ref{figCrystalGrowthRate}) and from this determine the melting temperature of the crystal as $T_{\rm melt}=0.447(2)$.

The composition of the crystal and the liquid was measured in terms of A and B species at $T = 0.42$. The crystal has a ratio of large to small particles of 2.35 which is close to the value of the 2:1 stoichiometry of the crystal, the over-representation of A particles is likely due to the the composition of the particles at the boundary between the crystal and the liquid being higher in A particles.

The LFS clusters in the liquid however have a ratio of 4.15 large particles to each small. This is close to the 4.23 expected for a 2:1 KA liquid, the composition of which is shown in Fig.~\ref{figKARatiosGraph}. This suggests that the structure of the LFS in the liquid is not strongly affected by the presence of the crystal throughout the simulation. Similarly, the percentage of particles in LFS clusters in the liquid region is about 10.5\% which is broadly in line with what is expected for a KA liquid at T = 0.42 as shown in Fig.~\ref{figAngell} b.

\section{Discussion}
\label{sectionDiscussion}

Changing the stoichiometry in the KA mixture leads to important changes in the relaxation dynamics of the mixture in the liquid phase. In particular, upon changing composition towards the 2:1 mixture, the system becomes more fragile. Richert and Angell~\cite{richert1998} suggested that more fragile glassformers should be expected to exhibit a greater change in structure upon supercooling, as evidenced by the jump in specific heat capacity observed in experiments on glassforming liquids upon falling out of equilibrium at $T_g$. While there are exceptions to this idea such as kinetically constrained models~\cite{pan2004}, spheres in higher dimension~\cite{charbonneau2013pre} and certain molecular glassformers~\cite{ito1999, ngai1999, martinez2001, huang2001}, many materials do follow this trend. In addition, recent work suggests a link between the size of correlated regions and fragility~\cite{bauer2013}, supporting the idea that glassformers in which the structure is more developed (involving more cooperation in the relaxation) are more fragile. In previous work on the Kob Andersen mixture and other model systems, the degree of structural change is correlated with fragility~\cite{royall2014, coslovich2007, coslovich2007ii} which is corroborated by recent work investigating the extent to which the structure predicts the dynamics~\cite{hocky2014, jack2014}.

With the above discussion in mind, our results which show the population of LFS decreasing upon changing the composition to the more fragile 2:1 mixture for a given supercooling [Fig. ~\ref{figAngell}(b) inset], seem odd at first sight. We cannot rule out the possibility of the formation of some other LFS, but we note the bicapped square antiprism would reasonably satisfy the criteria for a locally favoured structure given that the Al$_2$Cu crystal is stable. Furthermore, the transition between the liquid and the crystal appears quite strongly first order or at least there must be a strong free energy barrier to nucleation as we never observe it.

Intuitively, one might expect the freezing to be weakly first order since the LFS matches the crystal structure. An example of this is in the 2d hard disc system which forms a hexatic crystal in a weakly first order process upon increasing the volume fraction~\cite{bernard2011,engel2013}. However, we find no evidence of homogenous nucleation in our systems and conclude that they must be more strongly first order than considerations from d=2 would suggest.

We believe that the explanation lies in the competition between different varieties of LFS, each of which contributes differently to the stability and lifetimes of the overall LFS population. As schematically illustrated in Fig.~\ref{figKAtypes}, the three main types of bicapped structures differ in the role played by A (large) and B (small) particles along the principal rotational axis of symmetry of the bicapped antriprism. The identity of the capping particles is crucial, the structures with more of the larger A particles have a larger radius of gyration. The radii of gyration of $\rm{A_{8}B_{3}}$ and $\rm{A_{9}B_{2}}$ clusters are 4\% and 8\% smaller than the $\rm{A_{10}B_{1}}$. Therefore, the volume per cluster (and so the volume accessible via rotations around the principal axis of symmetry) is also larger for the $\mathrm{A_{10}B_{1}}$ and $\rm{A_{9}B_{2}}$ clusters than the $\rm{A_{8}B_{3}}$ cluster. The increase in accessible volume and the abundance of A particles with respect to B particles in all the considered mixtures, make the two energetically more costly structures (Fig.~\ref{figKARatiosGraph}b) more favourable from an entropic point of view. This behaviour is similar to that observed for a different set of competing clusters in the Wahnstr\"om mixture, an alternative model of colloidal suspensions~\cite{malins2009}. Together, the entropic contributions and the ground state energies of the clusters, make nucleation hardly accessible on one hand and on the other, seeded crystallisation possible.

We propose, that the $\mathrm{A_{8}B_{3}}$ LFS form a subset of LFS and that these may play a key role in slowing down the dynamics. Figure~\ref{figAngell}(b) inset implies that the 2:1 mixture exhibits a stronger dynamic response to the change in structure and this composition has a higher population of $\mathrm{A_{8}B_{3}}$ LFS. We therefore speculate that to understand the role of LFS, one should also consider not only its geometry but also its composition. If we assume that the $\mathrm{A_{8}B_{3}}$ is the key LFS, then the original hypothesis is satisfied in the sense that the ``domains'' of crystallisable LFS are larger in the 2:1 mixture than in the 4:1. Despite the fact that nucleation is not accessible in our computational time-scales, we expect that were the 2:1 mixture to be more deeply supercooled, then the original hypothesis might be satisfied, that at very deep supercooling, large domains of $\mathrm{A_{8}B_{3}}$ LFS would form. Whether such a deeply supercooled state point is sufficiently stable to crystallisation to observe these domains of LFS is open to question but in any case lies outwit our computational resources.

Any state with sizeable LFS domains would be highly supersaturated. Given our observation of crystal growth at moderate supercooling, it thus seems likely that were any large domains to form the system would then crystallise. We therefore conclude that the relief of frustration in the KA model by changing its composition may be affected by the need to supercool very deeply to generate large domains of ``crystalline'' LFS so the system would transition to a crystal. In this respect, it may be that the KA model satisfies the concept of geometric frustration via the change of composition rather than by curving space as is usually invoked.

Frustration has been successfully tuned in some d=2 systems by curving space~\cite{modes2007, sausset2008, sausset2010, tarjus2010, sausset2010pre} and other methods~\cite{shintani2006, kawasaki2007}. In d=2 the transition to crystallisation is much more weakly first order than seems to be the case here, so the LFS population in liquid can be much higher. Conversely, in d=3, control of frustration with an LFS compatible with the crystal, has been suggested in hard spheres and related systems~\cite{tanaka2010, kawasaki2010jpcm, leocmach2012}, but the prevalence of fivefold symmetric structures may complicate the situation~\cite{leocmach2012,royall2014}.

Finally we note that in certain lattice models more frustration against certain types of crystallisation can lead to a more complex crystal structure and more weakly first order freezing transition. An intriguing avenue for future research is how such observations translate in atomistic models such as that considered here~\cite{ronceray2015}.

\section{Conclusion}
\label{sectionConclusion}

In summary we have investigated the nature of geometric frustration in the Kob-Andersen model. Inspired by the observation that the 2:1 stoichiometry is compatible with the AlCl$_2$ crystal whose unit cell is consistent with locally favoured structure of the KA model (the bicapped square antiprism), we have explored the relief of geometric frustration via changing composition rather than by curving space. At first sight our results are surprising, rather than a strong increase in LFS population upon moving to the 2:1 composition, we find a small decrease. This is made more surprising by the observation that the 2:1 mixture is considerably more fragile than the 4:1 mixture, which is sometimes correlated with a relationship between structure and dynamics in the supercooled liquid.

We rationalise our observations by considering that the LFS themselves can play a role in the behaviour of the system. In particular, we have shown that for a given LFS population, those LFS whose composition is compatible with that of the Al$_2$Cu crystal, i.e. $\mathrm{A_{8}B_{3}}$ are correlated with slower dynamics. The population of these crystal-compatible LFS indeed increases markedly when the composition is changed to the 2:1 mixture, but remains a fraction of the total LFS population. Thus we do not find a large increase in LFS domain size upon changing the composition. Given that the LFS population overall increases upon supercooling, we speculate that a more deeply supercooled KA mixture than our simulations permit, may yet behave as expected.

\begin{acknowledgments}
This work was carried out using the computational facilities of the Advanced Computing Research Centre, University of Bristol. The open source molecular dynamics package LAMMPS~\cite{Plimpton1995} was used for molecular dynamics simulations. CPR acknowledges the Royal Society and all authors thank the European Research Council (ERC Consolidator Grant NANOPRS, project number 617266) for financial support
\end{acknowledgments}

\bibliographystyle{myaps}
\bibliography{frustration_arxiv}
\end{document}